\documentclass[10pt, onecolumn]{article}

\usepackage{bm,color,bbm}
\usepackage{ulem} %for strike-through font, e.g \sout{Bush} Obama is the president.
\usepackage{hyperref, mathtools,graphicx} %COMMENTED OUT BY MR!!!
\usepackage{mathtools,graphicx}
\usepackage{amssymb}

\newcommand{\beq}{\begin{equation}}
\newcommand{\eeq}{\end{equation}}
\newcommand{\bqa}{\begin{eqnarray}}
\newcommand{\eqa}{\end{eqnarray}}
\newcommand{\nn}{\nonumber}

\newcommand{\smallfrac}[2]{\mbox{$\frac{#1}{#2}$}}

\newcommand{\half}{\smallfrac{1}{2}}

\newcommand{\sq}[1]{\left[ {#1} \right]}

\newcommand{\tr}[1]{{\rm Tr}\sq{ {#1} }}

\newcommand{\id}{\mathbbm{1}}
\newcommand{\h}{{\cal H}}
\newcommand{\ce}{{\cal E}}
\newcommand{\cg}{{\cal G}}
\newcommand{\R}{{\cal R}}

\newcommand{\blk}{\color{black}}

\definecolor{maroon}{rgb}{0.7,0,0}

\definecolor{ngreen}{rgb}{0.3,0.7,0.3}

\definecolor{golden}{rgb}{0.8,0.6,0.1}

\begin{document}
\title{Entropic Heisenberg limits and uncertainty relations from the
	 Holevo information bound}
\author{
	Michael J. W. Hall\thanks{Department of Theoretical Physics, Research School of Physics and Engineering,
		Australian National University, Canberra ACT 0200, Australia}
	\thanks{Centre for Quantum Dynamics, Griffith University, Brisbane QLD 4111, Australia}}
\date{}
\maketitle

% USE THE FOLLOWING INSTEAD OF ABOVE FOR REVTEX (SEE ALSO PACS BELOW)
%\documentclass[preprint, showpacs, aps, draft]{revtex4}
%\usepackage{bm}

%\begin{document}
%\title{Prior information: how to beat the standard joint-measurement
%uncertainty relation}
%\author{Michael J. W. Hall}
%\affiliation{Theoretical Physics, IAS, \\ Australian National
%University,\\
%Canberra ACT 0200, Australia}
%\date{}

\begin{abstract}
Strong and general entropic and geometric Heisenberg limits are obtained, for estimates of multiparameter unitary displacements in quantum metrology, such as the estimation of a magnetic field from the induced rotation of a probe state in three dimensions. A key ingredient is the Holevo bound on the Shannon mutual information of a quantum communication channel. This leads to a Bayesian bound on performance, in terms of the prior distribution of the displacement and the asymmetry of the input probe state with respect to the displacement group. A geometric measure of performance related to entropy is proposed for general parameter estimation. It is also shown how strong entropic uncertainty relations for mutually unbiased observables, such as number and phase, position and momentum, energy and time, and orthogonal spin-1/2 directions, can be obtained from elementary applications of Holevo's bound. A geometric interpretation of results is emphasised, in terms of the `volumes' of quantum and classical statistical ensembles.
\end{abstract}

%USE FOR REVTEX
%\pacs{03.65.Ta}
%\maketitle

\section{Introduction}

Metrology is all about gaining information about one or more parameters of a physical environment or system, via interaction with a probe state. It would therefore seem natural to characterise the performance of a given metrology setup in information-theoretic terms.  Surprisingly, however, this is rarely considered thus far in quantum metrology (with some exceptions~\cite{yuen1, yuen2,  hallpra97,hallpra,spekkens,hallnjp,nair,hallprx12,sek,maccone}).  Attention has instead been largely focused on bounds for estimation uncertainties based on the quantum Cramer-Rao bound~\cite{helstrom, holevo, brauncaves,  glmreview,dattareview}. It should be noted that while the latter is formulated in terms of `Fisher information'~\cite{fisher}, this terminology predates that of information theory as introduced by Shannon~\cite{shan48}, and is not directly connected to entropy, coding or communication. Some comparisons between metrology bounds deriving from Shannon information and Fisher information may be found in~\cite{hallprx12,hallheislong,halltamagawa} (see also section~\ref{secentang}). It is the former that will be the focus of interest for this paper.

In classical metrology there is, in principle, no limitation on information gain. One can prepare and subsequently measure the configuration of a classical probe state precisely, and so estimate interaction parameters with arbitrary accuracy. In contrast, in quantum metrology the possible probe states after the interaction will typically overlap, even when pure, and hence cannot be perfectly resolved to obtain precise information about the interaction parameters. 
For example, the overlap between two phase-shifted optical modes $|\psi\rangle$ and $e^{-iN\theta}|\psi\rangle$ is $|\langle\psi|e^{-iN\theta}|\psi\rangle|^2= 1 -\theta^2(\Delta N)^2+O(\theta^4)$, implying that a small overlap, as required to resolve a small phase shift~$\theta$, requires a probe state with a correspondingly large photon number uncertainty $\Delta N\gtrsim 1/\theta$. 

A large photon number uncertainty does not in itself, however, place any clear restriction on the physical resources needed to achieve a given degree of phase resolution.  For example, the optical probe state with photon number decomposition $|\psi\rangle=(\sqrt{3}/2)\sum_n 2^{-n}|2^n\rangle$ has a divergent uncertainty $\Delta N$ and a relatively small average photon number $\langle N\rangle=3/2$~\cite{hallzz}, and yet, as shown below, is not useful for phase estimation.  Quantum bounds on metrological resolution in terms of physical resources, such as average photon number, entropy, number of modes, Hilbert space dimension, or interaction time, are called Heisenberg limits~\cite{glmreview,burnett,ou,zwierz}.  For example, it may be shown  that the root mean square error $\epsilon$, for any estimate $\theta_{est}$ of a random optical phase shift $\theta$, is bounded from below by the Heisenberg limits~\cite{hallpra}
\beq \label{heisrandom}
\epsilon:=\langle(\theta_{est}-\theta)^2\rangle^{1/2} \,> \,\sqrt{2\pi/e} \,\, 2^{-H(N)} >\frac{\sqrt{2\pi/e^3}}{\langle N\rangle + 1} ,
\eeq
where $H(N)$ denotes the Shannon entropy of photon number for the probe state. In particular, the root mean square error cannot scale better than inversely with $\langle N\rangle$, implying that the probe state in the example above has a rather poor phase resolution despite its divergent photon number uncertainty.

The first main aim of this paper is to obtain Heisenberg limits in the more general scenario of multiparameter estimation, where the probe state undergoes a unitary displacement $U_g$ parameterised by a group element $g\in{\cal G}$ for some (compact) group ${\cal G}$. Optical phase shifts as above correspond to the single parameter case, with $\cg\equiv U(1)$. A true multiparameter example of interest is the case of rotations in three dimensions, for example as induced by a magnetic field, with noncommuting group $\cg\equiv SO(3)$. Corresponding Heisenberg limits will be obtained for this case  in terms of the angular momentum properties of the probe state. 

In finding Heisenberg limits for general groups, an interesting issue arises with respect how the performance of an estimate should be characterised. For optical phase one can simply consider the statistics of the error in the estimate, $\theta_{err}:=\theta_{est}-\theta$, as above, and similarly for magnetic fields. However, more generally one cannot define a root mean square error for group elements, and a different approach is needed. The issue is resolved by showing that the entropy of the `control error',  $g_{err}:=g_{est}^{-1}\,g$, and the exponential thereof,  are suitable performance measures for the general case. It is shown that the maximum reduction in uncertainty of the displacement, as quantified by the change in entropy, is bounded by the asymmetry of the probe state with respect to the displacement group~\cite{spekkens,vaccaro1}, generalising results in~\cite{hallprx12} for phase shifts.  

The above results rely on the Holevo bound for the Shannon information of a quantum communication channel~\cite{hol}, and generalise its previous application to single parameter estimation~\cite{nair,hallprx12,sek,maccone}.  The second main aim of the paper is to show that the Holevo information bound further leads to a related simple proof of a strong entropic uncertainty relation for mutually unbiased (or complementary) quantum observables, where a sharp value of one observable implies a uniform distribution of the other. 

In particular, mutual unbiasedness of two observables $A$ and $B$, with eigenkets $\{|a\rangle\}$ and $\{|b\rangle\}$ respectively, corresponds to the property $|\langle a|b\rangle|^2=1/C_{AB}$ for all $a, b$, for some constant $C_{AB}$~\cite{kraus,colesreview}. Examples include position and momentum, number and phase, and orthogonal qubit observables. For any two such observables one has the entropic uncertainty relation~\cite{colbeck,frank}
\beq \label{hirgen}
H(A|\rho) + H(B|\rho) \geq S(\rho) + \log_2 C_{AB},\qquad C_{AB}\equiv  |\langle a|b\rangle|^{-2} ,
\eeq
significantly generalising an earlier result by Hirschman~\cite{hir}. Here $H(A|\rho)$ denotes the Shannon entropy of observable $A$ for a quantum system described by density operator $\rho$, and $S(\rho)$ denotes the von Neumann entropy of $\rho$ (explicit definitions are given in section~\ref{sec:hol}). Such entropic uncertainty relations place strong restrictions on the degree to which sharp values of observables can be simultaneously encoded in quantum states. They are  much stronger than variance-based uncertainty relations, and have found applications in many domains~\cite{colesreview}. 
%, and quantifies the inherent uncertainty or mixedness arising from the quantum state {\it per se}. 

Proofs of inequality~(\ref{hirgen}) in the literature, for various pairs of observables, are intrinsically difficult, as both classical and quantum entropies are involved~\cite{hallnjp,colbeck,frank,coles,kor,frank2}. For a Hilbert space having finite dimension $d$ it follows from a deep result for quantum conditional entropies~\cite{colbeck,coles}, or alternatively via the monotonicity of quantum relative entropy~\cite{kor}. For infinite dimensional Hilbert spaces it was proved by Frank and Lieb via the Golden-Thompson inequality for $C^*$-algebras~\cite{frank}.  

 In this paper it is shown that the strong entropic uncertainty relation~(\ref{hirgen}) can be obtained directly from the Holevo information bound. For example, for number and phase observables, the relation follows from the Holevo bound for the information that may be communicated via a uniform ensemble of phase-shifted signal states $\{ e^{-iN\theta}\rho e^{-iN\theta}\}$. More generally, it will be shown  for some specific examples \blk that equation~(\ref{hirgen}) corresponds to the Holevo bound for the case that observable $A$ is measured on a uniform ensemble of states that are related by unitary transformations generated by $B$.  This approach also leads to natural generalisations of the inequality for degenerate observables and for energy and time observables (with entropy replaced by almost-periodic entropy for the case of the time observable of a system with discrete energy eigenvalues~\cite{halltime}).

Finally, a third aim of the paper is to show how the results can be expressed in a natural geometric form, based on the fact that the exponential of the entropy of an ensemble, whether quantum or classical, is a direct measure of the effective volume or spread of the ensemble~\cite{hallvol}.  For example, the entropic uncertainty relation~(\ref{hirgen}) for the position $Q$ and momentum $P$ of a spin-zero particle moving in $D$ dimensions, with $C_{QP}=|\langle q|p\rangle|^{-2}=(2\pi\hbar)^D$,  can be rewritten as
\beq \label{xpgeo}
\frac{V_Q V_P}{h^D} \geq V(\rho) .
\eeq
Here $V_Q=2^{H(Q|\rho)}$ and $V_P=2^{H(P|\rho)}$ are the effective volumes of the position and momentum probability densities, $h:=2\pi\hbar$ is Planck's constant, and $V(\rho)=2^{S(\rho)}$ is the volume of the quantum state. It follows that {\it the minimum number of Planck volumes occupied by the ensemble, as quantified by the left hand side, is bounded by the Hilbert space volume occupied by the ensemble} (see also section~\ref{geom} below).  
A similar geometric interpretation applies to Holevo's information bound, analogous to a result by Shannon for classical Gaussian signals~\cite{shan49}, and to the various Heisenberg limits derived in the paper.
	
In section~\ref{sec:hol} the key ingredient required for the paper, i.e., the Holevo information bound, is briefly reviewed.  It is used to derive Heisenberg limits for multiparameter displacements in section~\ref{heislim}, and entropic uncertainty relations in section~\ref{eurs}. Conclusions are given in section~\ref{conc}.

\section{Entropy and the Holevo information bound}
\label{sec:hol}

The Holevo information bound places a limit on the degree of classical information that can be encoded into any ensemble of quantum states. It was first rigorously proved by Holevo for the case of finite set of signal states on a finite-dimensional Hilbert space~\cite{hol}, and extended to the general case by Yuen and Ozawa~\cite{yuen} (see Ref.~\cite{cd} for a brief history). The bound, and its derivation as a consequence of the decreasing distinguishability of statistical ensembles under measurement, are briefly reviewed in subsection~\ref{holdef}.  A geometric interpretation of the bound in terms of simple volume ratios,  arising from a related geometric interpretation of entropy, is reviewed in subsection~\ref{geom}.

\subsection{Decreasing distinguishability vs information gain}
\label{holdef}

Let $x$ label a set of quantum signal states $\{\rho_x\}$ incident on a receiver, where the state described by density operator $\rho_x$ is transmitted with prior probability density $p(x)$.  The label $x$  corresponds to the values of some random variable $X$, and \blk may be discrete or continuous. The ensemble of signal states will be denoted by ${\cal E}=\{\rho_x;p(x)\}$. Information may be recovered by measuring  some observable $A$ on each signal, yielding a corresponding maximum amount of error-free information gain per signal,  in the asymptotic limit, \blk of~\cite{shan48}
\beq \label{mutinf}
H(A:X) := H(A|\rho_{\cal E}) -\int dx\,p(x) H(A|\rho_x),
\eeq
known as the Shannon mutual information.  Here 
\beq \label{hent}
H(A|\rho):=-\int da\, p(a|\rho) \log_2 p(a|\rho)
\eeq
 denotes the Shannon entropy of the measurement distribution, $p(a|\rho):=\tr{A_a\rho}$, where $\{A_a\}$ is the positive operator valued measure (POVM) corresponding to $A$, and
 \beq 
 \rho_{\cal E} :=\int dx\, p(x)\rho_x
 \eeq
is the ensemble density operator.  Integration is replaced by summation in the above expressions for any discrete ranges of $x$ and $a$.  The choice of logarithm base in equation~(\ref{hent}) corresponds to a choice of units, with base 2 corresponding to the number of binary digits (bits) required to represent the information.
 
 The Holevo information bound for the Shannon mutual information is~\cite{hol,yuen}
 \beq \label{chi}
 H(A:X) \leq \chi({\cal E}) := S(\rho_\ce) - \int dx\,p(x) S(\rho_x),
 \eeq
 where 
 \beq \label{qent}
 S(\rho):=-\tr{\rho\log_2 \rho}
 \eeq 
 denotes the von Neumann entropy of state $\rho$. The bound thus limits the classical error-free information that can be encoded per signal of a given ensemble. It is saturable if and only if the density operators $\rho_x$ are mutually commuting.  However, if one allows a joint measurement on arbitrarily long sequences of signal states, rather than individual measurements on each signal state as above, then the bound $\chi(\ce)$ on information gain still applies, and is asymptotically saturable for any ensemble $\ce$ of signal states~\cite{hsw1,hsw2}.
 
 The general Holevo bound (including the cases of continuous labels $x$ and/or $a$, and infinite-dimensional Hilbert spaces), may be derived via the monotonicity of relative entropy under completely positive trace preserving mappings between states of $C^*$-algebras~\cite{yuen,uhl}. In particular, the mapping of a density operator to a classical probability density, via $\rho\rightarrow p(a)=\tr{\rho A_a}$, is such a mapping, thus yielding
 \beq \label{dd}
 H(p_1\| p_2) \leq S(\rho_1\|\rho_2) .
 \eeq
Here $H(p_1\|p_2):= \int da\, p_1(a)[ \log_2 p_1(a) - \log_2 p_2(a)]$ is the classical relative entropy of the measurement distributions for $\rho_1$ and $\rho_2$, and $S(\rho_1\|\rho_2) := {\rm Tr}[\rho_1 $ $(\log_2 \rho_1 - \log_2\rho_2)]$ is the quantum relative entropy of $\rho_1$ and $\rho_2$. In both cases the relative entropy is a nonnegative measure of distinguishability, vanishing only for the cases $p_1=p_2$ and $\rho_1=\rho_2$ respectively~\cite{disting}.  Thus, the inequality implies that {\it the distinguishability of two quantum states is at least as large as the distinguishability of their corresponding measurement distributions}, for any observable $A$.
The Holevo bound now follows via~\cite{yuen}
\beq \label{hol}
H(A:X) = \int dx\,p(x) H(p_x\| p_\ce) \leq \int dx p(x) S(\rho_x\|\rho_\ce)  = \chi(\ce),
\eeq
using definitions~(\ref{mutinf}) and~(\ref{chi}) and the decreasing distinguishability property~(\ref{dd}).

\subsection{Geometric interpretation of entropy and information}
\label{geom}
 
 %\subsubsection{Entropy and volume}
The above derivation of the Holevo bound relies on the decreasing distinguishability property in equation~(\ref{dd}), which is a relatively deep mathematical result~\cite{uhl}. Here an alternative approach is noted, based on the simple concept of the `volume' occupied by an ensemble.  An approach of this type was first given by Shannon in 1949, to bound the information content of classical Gaussian signals~\cite{shan49} (based on volumes of high-dimensional spheres), and was generalised to arbitary classical and quantum ensembles of signals 50 years later~\cite{hallvol}. It is encapsulated in the idea that error-free information must be coded in non-overlapping signals, corresponding geometrically to
\beq \label{wordeq}
{\rm information}~ = \log_2 (\#~ {\rm distinguishable~signals})\leq \log_2 \frac{\rm total ~signal ~volume}{\rm typical~ signal ~volume} .
\eeq
 The interested reader is referred to Ref.~\cite{hallvol} for details; here only the nature of the general link between entropy and volume will be considered, as it provides a useful intuition for understanding the entropic Heisenberg limits to be derived in section~\ref{heislim}. \blk

In particular, it is seen that the above approach  requires
a measure of the volume (or spread) of classical and quantum ensembles. Natural requirements for such a measure are~\cite{hallvol}:

\begin{description}
	\item[(i)~~] the volume of any mixture of non-overlapping ensembles, each of equal volume, is no greater than the sum of the component volumes (with equality for an equally-weighted mixture);

\item[(ii)~] the volume of an ensemble comprising two subsystems is no greater
than the product of the volumes of the subsystems (with equality when the
subsystems are uncorrelated); and

\item[(iii)] the volume of an ensemble is invariant under measure-preserving
transformations of the state space.
\end{description}
The above requirements are clearly independent of whether the ensemble is
classical or quantum, and are discussed in detail in Ref.~\cite{hallvol} (where the second postulate is shown to correspond to a projection property of Euclidean geometry: the product
of the lengths obtained by projecting a volume onto orthogonal axes is never
less than the original volume). 

Remarkably, the only continuous
measure of volume $V$ which satisfies the above postulates is
\beq \label{vol}
V = K\, 2^S, 
\eeq
where $K$ is a  positive \blk multiplicative constant, and $S$ denotes the Shannon entropy in equation~(\ref{hent}) for
classical ensembles (whether discrete or continuous), and the von Neumann entropy in equation~(\ref{qent}) for quantum ensembles. The constant $K$ corresponds to a choice of units, with a natural choice being $K=1$. 

In particular, choosing $K=1$, the volume is a measure of the effective number of bins over which a discrete probability density is spread, of the effective volume of the continuous space over which a continuous probability density is spread, and of the effective number of Hilbert space dimensions over which a quantum state is spread. It is minimised by pure states, and maximised by uniform states. For example, a discrete classical probability distribution $\{p_j\}$ over $d$ outcomes, and a density operator $\rho$ on a $d$-dimensional Hilbert space, each have have volumes $1\leq V\leq d$, with the upper bound corresponding to $p_j\equiv d^{-1}$ and $\rho=d^{-1}\id$ respectively.  Similarly, a continuous classical probability distribution uniform over some volume $v$, i.e., $p(x)=v^{-1}$, has volume $V=v$.

In this geometric approach to entropy, the volume $V$ of an ensemble can be taken as 
the primary physical quantity, with the entropy subsequently defined (up to an additive constant)
via 
\beq 
{\rm entropy} = \log_2 {\rm volume} .  
\eeq
This approach contrasts markedly with earlier axiomatic approaches by Shannon and others~\cite{shan48,behara}, where the latter apply only to discrete ensembles and lead to an arbitrary multiplicative constant for entropy rather than an additive constant (see~\cite{hallvol,hallvolcomm} for further discussion).  Note for the continuous case that a volume $V<1$ corresponds to a negative entropy. \blk

This connection between entropy and volume also yields a natural geometric interpretation of entropic uncertainty relations~\cite{hallvol}.  For example, as noted in equation~(\ref{xpgeo}) of the introduction, the entropic uncertainty relation for the position and momentum of a spin-zero particle moving in $D$ dimensions can be rewritten geometrically (choosing $K=1$ in equation~(\ref{vol})), as
\beq \nn
\frac{V_Q V_P}{h^D} \geq V(\rho) ,
\eeq
where $V_Q$ and $V_P$ are the ensemble volumes of the position and momentum distributions. Indeed, it was in this form that the strong entropic uncertainty relation for position and momentum was first conjectured to hold, via a semiclassical geometric argument~\cite{hallvol}. It was also shown there that the Boltzmann and Gibbs formulas for thermodynamic entropy, $S_B=k\log_2 W$ and $S_G=k S$,  where $k$ is Boltzmann's constant, \blk become equivalent when $W$ is interpreted as the number of zero-temperature volumes occupied by a thermal ensemble.
It will be seen in section~\ref{heislim} that entropic Heisenberg limits can also be rewritten in an intuitive geometric form.

\section{Deriving entropic and geometric Heisenberg limits}
\label{heislim}

The Holevo bound has  previously been successfully applied to obtain Heisenberg limits for phase estimation~\cite{nair,hallprx12}. This is  reviewed and generalised in section~\ref{phase}, and  further generalised to multiparameter estimation in section~\ref{multi}. An example of estimation of rotations in three dimensions, such as those induced by a magnetic field, is considered in section~\ref{magsec}.

\subsection{Phase estimation}
\label{phase}

The photon number operator $N$ of a single mode optical field generates a phase shift $\theta$ on a probe state $\rho$ via $\rho\rightarrow\rho_\theta:=e^{-iN\theta}\rho e^{iN\theta}$. The probe state can incorporate other degrees of freedom, in addition to the single mode acted on by $N$ (e.g., other optical modes with which it may be entangled). It is thus in general defined on some Hilbert space $\h=\h_{mode}\otimes \h'$ where $\h_{mode}$ is the Hilbert space of the mode. Typically, such a phase shift is generated by passing the mode through a medium, and estimation of the phase shift is used to gain information about properties such as the refractive index, path length or temperature of the medium. 

\subsubsection{Bounds on information gain}

If the prior probability density of the phase shift is denoted by $p_0(\theta)$, then the ensemble $\ce=\{\rho_\theta;p_0(\theta)\}$ describes the possible probe states following the phase shift.  It follows from the Holevo information bound~(\ref{chi}) that if the phase shift is estimated  via measurement of some observable $\Theta_{est}$ on the probe state, then the average information gained per probe state is bounded from above by
\beq \label{holphase}
H(\Theta_{est}:\Theta) \leq S(\rho_\ce) - \int d\theta\,p_0(\theta) S(\rho_\theta) = S(\rho_\ce) - S(\rho) .
\eeq
 The final equality follows noting that the von Neumann entropy is invariant under unitary transformations. 

To obtain an upper bound independent of the prior density $p_0(\theta)$, first define the phase randomisation operation $\R_\Phi$ via the unital map~\cite{spekkens, nair,vaccaro1}
\beq \label{rphi}
\rho_\Phi\equiv\R_\Phi(\rho):=\frac{1}{2\pi} \int d\phi\,e^{-iN\phi}\rho e^{iN\phi} = \sum_n \Pi_n \rho \Pi_n, 
\eeq
where  the factor $1/(2\pi)$ corresponds to a random phase shift $\phi$, and \blk $\Pi_n$ is the projection onto the eigenspace of the $n$th eigenvalue of $N$ (note this will be degenerate for $\h\neq\h_{mode}$). It follows that
\begin{align} \label{phirand}
\R_\Phi(\rho_\ce) =\int d\theta\,p_0(\theta)\, \frac{1}{2\pi} \int d\phi\, e^{-iN(\phi+\theta)}\rho e^{iN(\phi+\theta)} = \rho_\Phi ,
\end{align}
with the last equality obtained by replacing the integration variable $\phi$ by $\phi-\theta$. The von Neumann entropy is  non-decreasing \blk under unital operations, yielding
$S(\rho_\ce) \leq S(\R_\Phi(\rho_\ce)) = S(\rho_\Phi)$, which substituted into Eq.~(\ref{holphase}) gives the bound~\cite{spekkens,nair,hallprx12}
\beq \label{infbound}
H(\Theta_{est}:\Theta) \leq S(\rho_\Phi) - S(\rho) =: A_\cg(\rho) 
\eeq
on information gain. Note that the entropy difference $A_\cg(\rho)$ is the Holevo bound for the uniform ensemble $\ce_\Phi=\{\rho_\phi;(2\pi)^{-1}\}$. It may be recognised as the $\cg$-asymmetry (or asymmetry) of the probe state with respect to the group $\cg$ of phase shifts~\cite{spekkens,vaccaro1} (see~\cite{hallprx12} for further discussion in the metrology context), and is also a generalised measure of the coherence of the probe state with respect to $N$~\cite{baum,shuming}. 

The $\cg$-asymmetry is itself bounded by the photon number entropy $H(N|\rho)$ of the probe state~\cite{hallprx12}.  In particular, if $\sigma=|\Psi\rangle\langle|\Psi|$ is a purification of $\rho$ on some Hilbert space $\h\otimes \h_a$, with $\rho={\rm tr}_a[\sigma]$, then $\sigma_\phi:=e^{-iN\phi}\sigma e^{iN\phi}$ is a purification of $\rho_\phi$ with $\rho_\phi={\rm tr}_a[\sigma_\phi]$. Hence, using the monotonicity of relative entropy and the purity of $\sigma_\phi$ gives (correcting the derivation in~\cite{hallprx12}):
\begin{align}
A_\cg(\rho)&=\frac{1}{2\pi}\int d\phi\, S(\rho_\phi\|\rho_\Phi) \leq \frac{1}{2\pi}\int d\phi\, S(\sigma_\phi\|\sigma_\Phi) 
= S(\sigma_\Phi) \nn\\
&= S(\sum_n \Pi_n\otimes \id_a |\Psi\rangle\langle\Psi|\Pi_n\otimes \id_a) 
= S(\sum_n p_n |\Psi_n\rangle\langle\Psi_n|) \nn\\
&=-\sum_n p_n\log_2 p_n 
=H(N|\rho) \label{agbound}
\end{align}
as desired, with $p_n:=\langle\Psi|\Pi_n\otimes \id_a|\Psi\rangle=\tr{\Pi_n\rho}$ and $|\Psi_n\rangle:=p_n^{-1/2}(\Pi_n\otimes\id_a)|\Psi\rangle$ (thus $\langle\Psi_n|\Psi_{n'}\rangle=\delta_{nn'}$). 
Note that these bounds can be improved when the probe state is restricted to a single-mode state, i.e., $\h=\h_{mode}$, as one then has~\cite{hallprx12}
\begin{align} \label{phasesingle}
	A_\cg(\rho) &= S(\sum_n\Pi_n\rho\Pi_n)-S(\rho) =H(N|\rho) + \sum_n p_n S(\Pi_n\rho \Pi_n/p_n)-S(\rho) \nn\\
	&= H(N|\rho)-S(\rho) 
\end{align}
(using the orthogonality of the states $\{\Pi_n\rho \Pi_n/p_n\}$).

More generally, combining Eqs.~(\ref{holphase}), (\ref{infbound}) and~(\ref{agbound})  yields the inequality chain 
\beq \label{infbound2}
H(\Theta_{est}:\Theta)\leq S(\rho_\ce)-S(\rho) \leq S(\rho_\Phi)-S(\rho) \leq  H(N|\rho).
\eeq
This provides \blk strong upper bounds for the maximum information that can be gained about an unknown phase shift, that depend only on the initial probe state $\rho$ and the generator $N$. They immediately imply, for example, that at most one bit of information can be extracted via a NOON state $2^{-1/2}(|n,0\rangle+|0,n\rangle)$, even for nonlinear phase estimation~\cite{hallprx12}. They can also be generalised to include the effects of noise~\cite{nair} and entanglement~\cite{maccone}.

\subsubsection{Entropic and geometric Heisenberg limits}
\label{ehl}

The error in an estimate $\theta_{est}$ of a phase shift $\theta$ is given by
\beq \label{error}
\theta_{err} := \theta_{est} - \theta .
\eeq
For a good estimate, the probability density of $\theta_{err}$ will be concentrated about some value $\theta_0$ (note that any systematic error, corresponding to $\theta_0\neq0$, can be corrected via calibration).  Suitable measures of concentration, and hence of the performance of the estimate, are  the root mean square error, $\epsilon=\langle (\theta_{err})^2\rangle^{1/2}$, and the entropy of the error, $H(\Theta_{err})$.  Noting that phases are one-dimensional the latter measure may be more directly compared with the former via the associated ensemble {\it length} (see section~\ref{geom})
\beq \label{eps}
L_{err} := 2^{H(\Theta_{err})} \leq (2\pi e)^{1/2} \epsilon .
\eeq
Thus $0\leq L_{err}\leq 2\pi$, with the lower and upper bounds corresponding to perfect and random estimates,  i.e., $p(\theta_{err})=\delta(\theta_{err})$ and $p(\theta_{err})=(2\pi)^{-1}$, respectively. \blk The inequality in Eq.~(\ref{eps}) follows from the well known property that entropy is maximised by a Gaussian distribution over the real line  for distributions with a given variance and mean. \blk Some generic advantages of ensemble length over root mean square error, as a measure of the spread of a probability density, are discussed in~\cite{hallvol}.

Strong bounds on $H(\Theta_{err})$ and $L_{err}$, and hence on $\epsilon$, may be obtained from the previous section in combination with the lower bound~\cite{hallprx12}
\begin{align}
H(\Theta_{est}:\Theta) &= H(\Theta) - H(\Theta|\Theta_{est})= H(\Theta) - H(\Theta-\Theta_{est}|\Theta_{est}) \nn\\ \label{lower}
&\geq H(\Theta) - H(\Theta_{err}) %= \log_2 \frac{L_0}{L_{err}} .
\end{align}
Here the inequality follows via $H(A|B)=H(A)-H(A:B)\leq H(A)=H(-A)$ for the conditional entropy of $A$ given $B$. Note that the inequality is saturated for any {\it covariant} phase estimate, i.e., with $p(\theta_{est}|\theta+\phi) = p(\theta_{est}-\phi|\theta)$. 

One immediately has from equations~(\ref{holphase}), (\ref{infbound2}) and~(\ref{lower}) the entropic Heisenberg limits~\cite{hallprx12}
\beq \label{entdiff}
H(\Theta) - H(\Theta_{err}) \leq A_\cg(\rho) \leq H(N|\rho) .
\eeq
The left hand side is the reduction in uncertainty, as quantified by entropy, due to a given estimate. In particular, the initial uncertainty of the phase shift is given by the entropy $H(\Theta)$ of the prior probability density $p_0(\theta)$, and the final uncertainty is given the entropy $H(\Theta_{err})$ of the error in the estimate. Thus, {\it the maximum reduction in uncertainty, as quantified by entropy, is bounded by the $\cg$-asymmetry and the photon number entropy of the probe state}.

One can rewrite these inequalities in terms of direct measures of uncertainty, using the close connection between entropy and volume in section~\ref{geom}), leading  via Eq.~(\ref{infbound2}) \blk to the geometric Heisenberg limits
\beq \label{phasefrac}
\frac{L_{err}}{L_0} \geq \frac{V(\rho)}{V(\rho_\ce)} \geq \frac{V(\rho)}{V(\rho_\Phi)} \geq \frac{1}{V_N} .
\eeq
Here $L_0:=2^{H(\Theta)}$ is the ensemble length of the prior probability density $p_0(\theta)$, and $V_N:=2^{H(N|\rho)}$ is the volume of the photon distribution of the probe state.  The last inequality implies that {\it the uncertainty in phase, as quantified by the ensemble length, cannot be reduced by a factor greater than the effective number of photon states spanned by the probe state}.  This is a rather nice geometric constraint on the performance of any scheme for estimating an unknown phase shift. 

The intermediate inequalities in equation(\ref{phasefrac}) have similar geometric interpretations in terms of simple volume ratios. Note for the case $\h=\h_{mode}$ that the final upper bound can be strengthened to $V(\rho)/V_N$, via equation~(\ref{phasesingle}). Further, maximising the photon number entropy for a fixed value of $\langle N\rangle$ yields
\beq \label{nav}
\frac{L_{err}}{L_0} >  \frac{1}{e(\langle N\rangle + 1)},
\eeq
and hence the reduction in uncertainty cannot scale better than linearly with the average photon number.

Equations~(\ref{eps}), (\ref{phasefrac}) and(\ref{nav}) may be combined to obtain corresponding lower bounds for the root mean square error $\epsilon$.  For example, the Heisenberg limits in equation~(\ref{heisrandom}) of the introduction are equivalent to~\cite{hallpra,hallnjp}
\beq \label{epsheis}
\epsilon > (2\pi e)^{-1/2} \frac{L_0}{V_N} > (2\pi e^3)^{-1/2} \frac{L_0}{\langle N\rangle + 1} ,
\eeq
for the choice $L_0=2\pi$ (corresponding to a random phase shift $p_0(\theta)=(2\pi)^{-1}$).

\subsubsection{Multimode phase estimation}
\label{secentang}
	
It is of interest to consider the case where $M$ modes, described by some possibly entangled probe state $\rho_M$, each undergo a phase shift $\theta$ and are measured jointly to estimate $\theta$. This corresponds to the unitary transformation $\rho_M\rightarrow e^{-iN_T\theta}\rho_M e^{-iN_T\theta}$ on the joint state, where $N_T:=N_1+N_2+\dots N_M$ denotes the total photon number operator. 
The previous derivations go through precisely as before, with $N$ replaced by $N_T$, as they only rely on the property of $N$ having integer eigenvalues (excepting equation~(\ref{phasesingle}), which further requires $N$ to be nondegenerate on $\h$).  In particular, equations~(\ref{phasefrac}) and~(\ref{nav}) lead to
\beq \label{mmode}
\frac{L_{err,M}}{L_{0,M}} \geq 2^{-H(N_T|\rho_M)} >  \frac{1}{e\,(\langle N_T\rangle + 1)}.
\eeq
for multimode probe states.

It is also of interest to examine the asymptotic scaling properties of this bound for large $M$.
First, for the general case,   note that defining the $M\times M$ covariance matrix $C\geq0$ by $C_{jk}:=\langle N_jN_k\rangle-\langle N_j\rangle\langle N_k\rangle$, one has
\[
{\rm Var} \,N_T =\sum_{j,k} C_{jk} \leq \sum_{j,k}\sqrt{C_{jj}C_{kk}}=(\sum_j 
\sqrt{C_{jj}})^2 =(\sum_j \Delta N_j)^2 .\]
Hence, \blk using the variational inequality 
\beq \label{mow}
H(G) \leq \half \log_2 2\pi e [{\rm Var}\, G+1/12]
\eeq
for the entropy and variance of an integer-valued random variable $G$~~\cite{mow},  equation~(\ref{mmode}) leads to the multimode entropic Heisenberg limit
\beq
\frac{L_{err,M}}{L_{0,M}} \geq \frac{1}{\sqrt{2\pi e[ M^2 (\overline{\Delta N})^2 +1/12]}} ,
\eeq
where $\overline{\Delta N}$ denotes the average root mean square error $M^{-1}\sum_j \Delta N_j$. Thus, the bound scales inversely with $M$ as $M\rightarrow\infty$.

In contrast, for a probe state comprising a product state of $M$ independent modes, $\rho_M=\otimes_j\rho_j$,  one has $C_{jk}=0$ for $j\neq k$, yielding ${\rm Var} \,N_T= \sum_j C_{jj} =\sum_j (\Delta N_j)^2$, \blk and leading to the bound
\beq \label{prod}
\frac{L_{err,M}}{L_{0,M}} \geq \frac{1}{\sqrt{2\pi e[ M \overline{(\Delta N)^2} +1/12]}} ,
\eeq 
 where $\overline{(\Delta N)^2}$ denotes the average variance $M^{-1}\sum_j{\rm Var}\, N_j$. This bound \blk scales inversely with $\sqrt{M}$, and  thus, entangled probe states can have up to a $\sqrt{M}$ advantage, consistent with other metrology bounds~\cite{maccone,glmreview,dattareview,hallzz}.

Finally,  for the case of $M$ identical copies of a single mode $\rho$, i.e, for $\rho_M=\otimes^M\rho$, equation~(\ref{mmode}) directly yields
 \beq
 \frac{L_{err,M}}{L_{0,M}} \geq 2^{-H(N_1+\dots N_M|\otimes^M\rho)}\approx 2^{-\half \log_2 2\pi e M(\Delta N)^2} = \frac{1}{\sqrt{2\pi e M}\,\Delta N}
\eeq
as $M\rightarrow\infty$, using the central limit theorem that the distribution of a sum of identically distributed independent variables approaches a Gaussian distribution having the same variance. Combining this result with equation~(\ref{eps}) gives the asymptotic scaling
\beq
\epsilon \gtrsim \frac{L_0}{2\pi}\,  \frac{1}{e\sqrt{M}\, \Delta N} \blk
\eeq
for the root mean square error, which is similar in form to the quantum Cramer-Rao bound $(2\sqrt{M}\,\Delta N)^{-1}$ for the measurement uncertainty in an unbiased estimate of phase~\cite{brauncaves, glmreview,dattareview}. The latter bound is asymptotically saturable, suggesting that the entropic and geometric Heisenberg limits are similarly saturable to within a scaling factor.

\subsection{Multiparameter group estimation}
\label{multi}

To generalise from phase estimation to multiparameter estimation, consider now unitary displacements of a quantum state indexed by some (compact) group $\cg$, $\{U_g:g\in\cg\}$, which form a projective representation of the group, i.e., 
\beq
U_gU_h\,\rho \,U_g^\dagger U_h^\dagger = U_{gh}\,\rho \,U_{gh}^\dagger
\eeq
for all $g,h\in\cg$ and probe states $\rho$. The phase estimation scenario corresponds to the single parameter group $\cg\equiv U(1)$.  In the more general scenario a probe state undergoes a unitary displacement corresponding to $g$ with prior probability $p_0(g)$, and the displacement is estimated via a subsequent measurement on the probe state. The aim of this subsection is to obtain corresponding bounds on the performance of the estimate, in the form of information bounds and entropic Heisenberg limits. The example of the three-dimensional rotation group is then discussed in the following subsection.

The first step, just as for the phase estimation scenario, is to use Holevo's information bound. Defining $\rho_g:=U_g\,\rho\,U_g^\dagger$, the ensemble of probe states following the displacement is $\ce=\{\rho_g;p_0(g)\}$, and the Holevo bound~(\ref{chi}) for the maximum error-free information gained per probe state, via a measurement of some observable $G_{est}$, is
\beq \label{ghol}
H(G_{est}:G) \leq S(\rho_\ce) - S(\rho) ,\qquad\qquad \rho_\ce=\int_\cg dg \,p_0(g) \rho_g,
\eeq
Here $dg$ denotes the (unique) invariant normalised right Haar measure for the group~\cite{haar}.  Defining the displacement randomisation operation $\R_\cg$ via the unital map~\cite{spekkens,vaccaro1}
\beq \label{rg}
\rho_\cg\equiv\R_\cg(\rho) := \int_\cg dg\,U_g\rho U_g^\dagger ,
\eeq
analogously to $\R_\Phi$ in equation~(\ref{rphi}), one has the generalisation
\beq
\R_\cg(\rho_\ce) = \int_\cg dh\,p_0(h)\int_\cg dg\,\rho_{gh} = \R_\cg(\rho)
\eeq
of equation~(\ref{phirand}), where the second equality follows by replacing the integration variable $g$ by $gh^{-1}$ and using $d(gg')=dg$ for any $g'\in\cg$). Hence, equation~(\ref{infbound}) generalises to the information bound~\cite{spekkens}
\beq \label{ginf}
H(G_{est}:G) \leq S(\rho_\cg) - S(\rho) = A_\cg(\rho) ,
\eeq
for multiparameter group estimation. Here $A_\cg(\rho)$ is the Holevo bound for the uniform ensemble $\ce_\cg=\{\rho_g; v^{-1}\}$, with $v:=\int_\cg dg$, and is similarly equal to the $\cg$-asymmetry of the probe state~\cite{spekkens,vaccaro1}, and to a generalised measure of its coherence~\cite{shuming}. 

Equation~(\ref{ginf}) bounds the maximum information that can be gained for a given displacement group and probe state. However, to obtain more direct bounds on the performance of the estimate, corresponding to Heisenberg limits, one needs a concept of the `error' in the estimate.  A natural candidate is the {\it control error} associated with an estimate $g_{est}$ of displacement $g$, defined by
\beq \label{gerr}
g_{err} := g_{est}^{-1}\,g ,
\eeq
in analogy to equation~(\ref{error}).  In particular, if one is using an estimate to control the state $\rho$, by correcting the actual displacement $g$ by applying $g_{est}^{-1}$ (e.g., undoing a phase shift or a magnetically-induced rotation of the state), then the displacement followed by the correction is equivalent to applying $g_{err}$ to the initial state.  

For a good estimate, the probability density of $g_{err}$ will be concentrated about some value $g_0$ (with any systematic error $g_{err}\neq e$ correctable via calibration, where $e$ is the identity element). 
For a general group  there is no natural measure of root mean square error to characterise this degree of concentration. However, the entropy $H(G_{err})$ and corresponding ensemble volume $V_{err}=2^{H(G_{err})}$ of $p(g_{err})$ are suitable measures of concentration, just as for the case of phase. In particular, $V_{err}$ is a direct measure of the effective group volume occupied by $p(g_{err})$ (see section~\ref{geom}). 

To obtain entropic Heisenberg limits for any estimate of $g$, note first that one has the lower bound
\begin{align}
H(G_{est}:G) &= H(G) - H(G|G_{est}) = H(G) - H(G_{est}^{-1}G|G_{est})\nn\\
&\geq H(G)-H(G_{err}) = \log_2 \frac{V_0}{V_{err}} , 
\end{align}
for the mutual information. Here $V_0:=2^{H(G)}$ is ensemble volume of the prior probability density $p_0(g)$, and the second equality follows using $p(g') dg'=p(hg) dg$ for $g'=hg$, since the left and right invariant Haar measures are equal for all compact groups~\cite{haar}. Combining this with the upper bounds~(\ref{ghol}) and~(\ref{ginf}) then yields the entropic Heisenberg limit
\beq \label{heisag}
 H(G)-H(G_{err}) \leq A_\cg(\rho) ,
\eeq
bounding the reduction in uncertainty by the $\cg$-asymmetry similarly to equation~(\ref{entdiff}) for phase.  One also has from these equations the geometric Heisenberg limits
\beq \label{genlim}
\frac{V_{err}}{V_0} \geq \frac{V(\rho)}{V(\rho_\ce)}\geq \frac{V(\rho)}{V(\rho_\cg)} =2^{-A_\cg(\rho)},
\eeq
Thus, {\it the uncertainty of the displacement, as quantified by ensemble volume, cannot be reduced by a factor greater than the exponential of the $\cg$-asymmetry of the probe state}. Noting that $V(\rho)\geq 1$, one also has the fundamental result
\beq \label{fundamental}
\rm final~volume~ of~ uncertainty \geq \frac{\rm initial~ volume ~of~uncertainty}{\rm volume~ of~ randomly~ displaced ~probe ~state} .
\eeq
These results generalise the phase estimation limits in equation~(\ref{phasefrac}) to all compact multiparameter groups.
It can also be shown that the above bounds are unaffected if one replaces $g_{err}$ by its inverse (essentially because $H(G_{est}:G)=H(G_{est}^{-1}:G)$), or by $g\,g_{est}^{-1}$. % (essentially because the left and right invariant Haar measures are equal for compact groups). 

\subsection{Example: rotations in 3 dimensions}
\label{magsec}

Consider now the rotation group $\cg\equiv SO(3)$. For a probe state with angular momentum operator $\bm J=(J_x,J_y,J_z)$,  a rotation about the unit direction $\bm {\hat n}$ by an angle $\theta$ corresponds to the unitary operator 
\beq  \label{un}
U_{\bm {n}} = e^{-i \bm J\cdot\bm n/\hbar},
\eeq
 with $\bm n:=\theta\bm {\hat n}$. The group is compact, with a rotation of $\theta$ about $\bm{\hat n}$ equivalent to a rotation of $2\pi-\theta$ about $-\bm{\hat n}$, so that one can restrict $\theta=|\bm n|$ to the interval $[0,\pi]$. 
For a spin-$j$ particle with magnetic moment $\mu$, in a constant magnetic field $\bm B$ for a fixed time $T$, one has $\bm n=\mu T\bm B$.  Hence, an estimate of $\bm n$ can be used to estimate any of $\mu$, $T$ or $\bm B$ when the other two are known.

\subsubsection{Estimating a rotation}

To apply the bounds in equation~(\ref{genlim}) to the estimation of $\bm n$, it is necessary to calculate the $\cg$-asymmetry $A_\cg(\rho)$ for the rotation group. This will be done here for the case where only the angular momentum degrees of freedom of the probe state are relevant, so that the relevant Hilbert space $\h$ is the span of the angular momentum eigenstates  in the usual $(J^2,J_z)$ basis, $\{|j,m\rangle\}$, with half-integer $j$ and $m=-j,-j+1,\dots, j-1,j$.  Hence, the randomisation of the probe state over all rotations can be written via equation~(\ref{rg}) as
$
\rho_\cg = \int d\bm n\, e^{-i \bm J\cdot\bm n/\hbar} \rho e^{i \bm J\cdot\bm n/\hbar} ,
$
where $d\bm n$ denotes the invariant Haar measure. Clearly, this state is invariant under any rotation, and hence must commute with $\bm J$ and have the same statistics as $\rho$ for $J^2$, leading uniquely to
\beq
\rho_\cg = \sum_j \tr{\Pi_j\rho} \frac{\Pi_j}{2j+1},\qquad\qquad \Pi_j:=\sum_{m=-j}^j |j,m\rangle\langle j,m| .
\eeq
Here $\Pi_j$ is the projection onto the $j$th eigenspace of $J^2$, with dimension $\tr{\Pi_j}=2j+1$.  It follows, defining  $p_j:=\tr{\Pi_j\rho}$, that the eigenvalues of $\rho_\cg$ are of the form $\lambda_j=p_j/(2j+1)$ with corresponding degeneracy $2j+1$. Hence, the von Neumann entropy of $\rho_\cg$ is
\beq
S(\rho_\cg)= - \sum_{j} \sum_{m=-j}^j \frac{p_j}{2j+1}\log_2 \frac{p_j}{2j+1} = H(J^2|\rho) + \sum_j p_j \log_2 (2j+1) ,
\eeq
and the $\cg$-asymmetry~(\ref{ginf}) of the probe state  follows  as
\beq \label{aggen}
 A_\cg(\rho) = H(J^2|\rho) + \langle \log_2(2j+1)\rangle - S(\rho).
 \eeq
For the case of a spin-$j$ particle one has the fixed value $J^2= j(j+1)\hbar^2$, yielding $H(J^2|\rho)=0$ and the simplification
\beq \label{agspin}
 A_\cg(\rho) = \log_2 (2j+1) - S(\rho) .
 \eeq

Equations~(\ref{aggen}) and~(\ref{agspin}) bound the information which can be gained by any estimate of a rotation in three dimensions, as per equation~(\ref{ginf}), in terms of the angular momentum properties of the probe state.  They may also be used to bound the degree to which an estimate can reduce the degree of uncertainty of a rotational displacement, via equation~(\ref{genlim}). For example, for a probe state comprising a spin-$j$ particle, equations~(\ref{genlim}) and~(\ref{agspin}) yield the corresponding geometric Heisenberg limit
\beq \label{heisrot}
\frac{V_{err}}{V_0} \geq \frac{V(\rho)}{2j+1} \geq \frac{1}{2j+1},
\eeq
where the second inequality follows immediately from the property $V(\rho)\geq1$ for any quantum state $\rho$.  Thus the uncertainty can be reduced by no more than a factor of $2j+1$ for this case.
In contrast, for a probe state with support only on values $j\leq j_{max}$, a simple variational calculation gives
\beq \label{agjmax}
A_\cg(\rho) \leq \log_2(j_{max}+1)^2-S(\rho),
\eeq
with equality for the case $p_j\propto 2j+1$, and hence allows for a greater potential reduction in uncertainty, corresponding to the Heisenberg limit
\beq \label{jmax}
\frac{V_{err}}{V_0} \geq \frac{V(\rho)}{(j_{max}+1)^2} \geq \frac{1}{(j_{max}+1)^2}  .
\eeq

\subsubsection{Estimating a magnetic field}

The rotational control error $\bm n_{err}$ corresponding to equation~(\ref{gerr}) is defined for a given estimate $\bm n_{est}$ of $\bm n$ via 
\beq
e^{-i \bm J\cdot\bm n_{err}/\hbar} = e^{i \bm J\cdot\bm n_{est}/\hbar} e^{-i \bm J\cdot\bm n/\hbar} ,
\eeq
from equation~(\ref{un}). For the case where one is estimating a magnetic field $\bm B$, related to $\bm n$ via $\bm n=\mu T\bm B$,  then $\bm B_{err}$ is the effective net magnetic field experienced when the estimate is used to correct for the rotation induced by $\bm B$ (see  discussion in section~\ref{multi}).

However, outside the control and tracking context, it is also natural to  consider the error defined by the difference between $\bm B$ and its estimate, i.e.,
\beq \label{berr}
\tilde {\bm B}_{err} := {\bm B}_{est} - {\bm B} .
\eeq
This is of interest in quantifying the accuracy to which the magnetic field  per se can be estimated, in contrast to the accuracy of the corresponding rotation.  It is not difficult to obtain corresponding entropic Heisenberg limits for this error, and to relate them to various measures of uncertainty.

First, since $\bm n$ and $\bm B$ are in one-one correspondence, one immediately has the mutual information bound
\beq \label{infmag}
H(\bm B_{est}:\bm B) = H(\bm n_{est}:\bm n) \leq A_\cg(\rho) = H(J^2|\rho) + \langle \log_2(2j+1)\rangle - S(\rho) ,
\eeq 
for any estimate of the magnetic field, using equations~(\ref{ginf}) and~(\ref{aggen}).  Second, a simple generalisation of the phase estimation lower bound in equation~(\ref{lower}) yields the lower bound
\begin{align}
H(\bm B_{est}:\bm B) &= H(\bm B) - H(\bm B|\bm B_{est})= H(\bm B) - H(\bm B-\bm B_{est}|\bm B_{est}) \nn\\ \label{lowerb}
&\geq H(\bm B) - H(\tilde{\bm B}_{err}) = \log_2 \frac{V_0(\bm B)}{V_{err}(\bm B)} 
\end{align}
for information gain, where $V_0(\bm B)=2^{H(\bm B)}$ denotes the  volume of uncertainty associated with the prior probability density $p_0(\bm B)$ of the magnetic field ($\bm B$ appears explicitly in the volumes above, to distinguish them from the volumes associated with estimates of the corresponding rotation). Combining the above two equations immediately gives the geometric Heisenberg limit
\beq \label{mag1}
\frac{V_{err}(\bm B)}{V_0(\bm B)} \geq 2^{-A_\cg(\rho)} =\frac{V(\rho)}{2^{\langle \log_2 (2j+1)\rangle}\, V_{J^2}}  .
	\eeq  
In particular, for a probe state consisting of a spin-$j$ particle this reduces to
\beq \label{mag2}
\frac{V_{err}(\bm B)}{V_0(\bm B)} \geq  \frac{V(\rho)}{2j+1} \geq \frac{1}{2j+1} ,
\eeq
similarly to equation~(\ref{heisrot}) for estimates of rotations per se.

There is an important subtlety here that should be noted:  the above results only apply to estimates of $\bm B$ modulo a multiple of the magnitude $B_{\pi}:=2\pi\mu T$. In particular, a rotation of the probe state by a field of magnitude $B_\pi$, about unit direction $\bm {\hat n}$,  is equivalent to a rotation of the same magnitude about $-\bm {\hat n}$,  and hence these cannot be physically distinguished.  Hence, all probability densities and volumes appearing in the above results are restricted to magnetic fields with $|\bm B|\leq B_{\pi}$.  This issue, that one can only estimate the magnitude up to a multiple of $B_{\pi}$ in the absence of additional information, is similar to the periodicity of optical phase estimates, but appears to have been neglected in the quantum metrology literature. \blk

A further and useful subtlety is that the derivation of the lower bound~(\ref{lowerb}) is independent of the measure used to define the probability density of the magnetic field (since mutual information is invariant under one-one transformations). Hence the bounds~(\ref{mag1}) and~(\ref{mag2}) in fact hold for any choice of measure, such as the Lebesgue measure on the space of magnetic fields  $|\bm B|\leq B_{\pi}$. For  the latter case if follows that  the maximum ensemble volume is  $4\pi(B_{\pi})^3/3$. 

One can also obtain entropic Heisenberg limits for other measures of error for magnetic fields, by relating these measures to the entropy. For example,  the error matrix $\tilde E$ 
\beq
\tilde E := \langle \tilde {\bm B}_{err} \tilde {\bm B}_{err}^\top\rangle ,
\eeq
for the error $\tilde {\bm B}_{err}$ in equation~(\ref{berr}) generates two associated measures of error, 
\beq \label{dbound}
D_{err} := (\det \tilde E)^{1/2}, \qquad\qquad T_{err}:= ({\rm tr}[\tilde E])^{1/2} = \langle |\bm B_{est}-\bm B|^2\rangle^{1/2}.
\eeq
These may be regarded as generalisations of the root mean square error in equation~(\ref{eps}) (one can also define a related matrix and measures for $\bm B_{err}$). Note that $D_{err}$ is a measure of volume in the three-dimensional space of magnetic fields, while $T_{err}$ is a measure of length. Choosing the Lebesgue measure as per the above paragraph, entropy is maximised for a fixed error matrix $\tilde E$ by a Gaussian distribution having the same error matrix and zero mean. Together with the relation $\det \tilde E \leq ({\rm tr}[\tilde E]/3)^3$ (following from comparing the arithmetic and geometric means of the eigenvalues), this leads to 
\beq
V_{err} \leq (2\pi e)^{3/2} D_{err} \leq (2\pi e/3)^{3/2} (T_{err})^3.
\eeq
Thus, using equation~(\ref{mag1}) gives the corresponding Heisenberg limits
\begin{align} \label{something}
D_{err} &\geq (2\pi e)^{-3/2} \,2^{H(\bm B)}\, 2^{-A_\cg(\rho)},
\\
 \label{terr}
T_{err} &\geq (2\pi e/3)^{-1/2}\,  2^{H(\bm B)/3}\, 2^{-A_\cg(\rho)/3} .
\end{align}  
For example, for the case of a spin-$j$ particle and a random prior distribution $p_0(\bm B)= (4\pi B_{\pi}^3/3)^{-1}$, equations~(\ref{agspin}) and~(\ref{terr}) yield the bound
\beq \label{brand}
T_{err} \geq \frac{(\pi e^3/6)^{-1/6} }{(2j+1)^{1/3}} \, B_{\pi} \approx \frac{0.6756}{(2j+1)^{1/3}} \, B_{\pi} .
\eeq

One can also consider the case of $M$ (possibly entangled) spin-$j$ particles, each undergoing the same rotation.  This corresponds to the unitary transformation $e^{-i\bm J_T\cdot\bm n}$ acting on a composite probe state $\rho_M$, with $\bm J_T:=\bm J_1+\bm J_2+\dots +\bm J_M$. The joint Hilbert space is thus a direct sum of spins ranging from 0 up to $Mj$, which corresponds to $j_{max}=Mj$ in equation~(\ref{agjmax}). Hence, for this case the corresponding bounds on resolution are approximated by replacing $2j+1$ by $(Mj+1)^2$ in the above results.  This gives, for example, the asymptotic scalings 
\beq
V_{err}(\bm B)\sim M^{-2},\qquad\qquad  T_{err}\sim M^{-2/3} \eeq
for the lower bounds. 

A related scenario of interest to consider is where the probe state comprises $M$ pairs of entangled spin-$j$ particles with only one member of each pair rotated by the field~\cite{datta,yuan}.  For example, for $M$ qubit pairs (i.e, $j=1$), with each pair described by a singlet state, it is easy to calculate the corresponding $\cg$-asymmetry for $M=1$ to be $\log_2 2$, which is equivalent to replacing the denominator $2j+1=2$ in the above results by $2^{\log_2 2}=2$, i.e, the bound for a singlet state in this scenario is the same as for a single qubit. Calculation of the $\cg$-asymmetry for $M>1$ singlets is more difficult, and left to future work. The asymptotic scaling of $T_{err}$ with $M$ is conjectured to be weaker than the corresponding Cramer-Rao bound for unbiased estimates (which scales as $M^{-1}$~\cite{datta,yuan}), leading to stronger bounds for large $M$. This would be analogous to the case of phase estimation with NOON states mentioned following equation~(\ref{phasesingle}): the Cramer-Rao bound for the root mean square error $\epsilon$ scales as $N^{-1}$ for this case, whereas the first Heisenberg limit in equation~(\ref{epsheis}) gives a constant bound independent of $N$ (see~\cite{hallprx12,halltamagawa} for further discussion).

Finally, it would be worthwhile to calculate the Heisenberg limits corresponding to estimation of the direction of an unknown rotation axis or magnetic field, for the case where the angle of rotation or the magnitude of the  field is known. Note that the converse problem, of estimating an unknown rotation angle or field magnitude for a known fixed direction, the $z$-direction say, is equivalent to considering the group of rotations corresponding to $U(1)\equiv\{ e^{-iJ_z\theta/\hbar}\}$. This is isomorphic to the group of phase shifts, and hence the results of section~\ref{phase} hold for this case, with $H(N|\rho)$ replaced by $H(J_z|\rho)$ and $\langle N\rangle$ by $2\hbar^{-1}\langle |J_z|\rangle$~\cite{hallpra,hallnjp}.

\section{Deriving entropic uncertainty relations}
\label{eurs}

Recall from the introduction that the mutual unbiasedness of two observables $A$ and $B$, with POVMs $\{|a\rangle\langle a|\}$ and $\{|b\rangle\langle b|\}$ respectively, corresponds to the property that a sharp value of one observable implies a uniform distribution of the other, with  $|\langle a|b\rangle|^2=1/C_{AB}$ for all $a, b$, for some constant $C_{AB}$~\cite{kraus,colesreview}. Examples include position and momentum, number and phase,  orthogonal qubit observables, and energy and time.  In 1957 Hirschman used norm inequalities for Fourier transform pairs to obtain the entropic uncertainty relation
\beq \label{hir}
H(A|\rho) + H(B|\rho) \geq \log_2 C_{AB}, \qquad\qquad C_{AB}\equiv  |\langle a|b\rangle|^{-2},
\eeq
for any pair of mutually unbiased observables $A$ and $B$~\cite{hir}.
Hirschman gave explicit examples of inequality~(\ref{hir}) for the case of one-dimensional position and momentum observables $Q$ and $P$, for which $C_{QP}=2\pi\hbar$, and for angular momentum and phase observables $J_z$ and $\Phi$, for which $C_{J_z\Phi}=2\pi$. The inequality was independently proved by Mamojka in 1974 for the case of mutually unbiased qubit observables $\sigma\cdot\bm  m$ and $\sigma\cdot\bm  n$, with $\bm  m\cdot \bm  n=0$ and $C_{\bm  m \bm  n}=2$, corresponding to orthogonal measurement directions on the Bloch sphere~\cite{mam}.  The case of energy and time observables appears to have first been considered by Grabowski~\cite{grab} for free particles and periodic systems, and later generalised to systems with arbitrary discrete energy spectra via almost-periodic entropies~\cite{halltime}.

Both Hirschman and Mamojka conjectured that the right hand side of inequality~(\ref{hir}) could be improved to $\log_2 e\pi\hbar$ for position and momentum observables, as was later verified by Beckner~\cite{beck} and in more detail by Bialynicki-Birula and Mycielski~\cite{bbm}. However, the inequality is tight whenever at least one observable is discrete-valued, being saturated by an eigenstate of the observable~\cite{hir}. Generalisations to state-independent bounds for pairs of non-mutually unbiased observables have been given by various authors~\cite{colesreview}, most notably Maassen and Uffink, who showed for arbitrary Hermitian observables on finite-dimensional Hilbert spaces that $C_{AB}$ can be replaced by $\max_{a,b} |\langle a|b\rangle|^{-2}$~\cite{maas} (thus proving a conjecture by Kraus~\cite{kraus}), and  Krishna and Parthasarathy, who generalised the Maassen-Uffink bound to arbitrary  POVMs~\cite{partha}.

In all of the above examples the uncertainties of  $A$ and $B$, as quantified by the entropies of their classical measurement distributions, are precluded from jointly approaching their minimum values by a state-independent lower bound.  However, it is clearly of interest to consider stronger bounds, that take into account the contribution of any inherent quantum uncertainty in the state itself.  For example, the thermal state of a harmonic oscillator becomes more mixed as the temperature increases.  This mixedness feeds into and increases the oscillator's position and momentum uncertainties, such that equation~(\ref{hir}) becomes trivial even for relatively low thermal energies $E= \hbar\omega \langle N +\half\rangle\geq \hbar\omega$~\cite{hallnoise}. 

Such state-dependent effects are taken into account by the highly nontrivial generalisation given in equation~(\ref{hirgen}) of the introduction  proved by Berta~{\it et al.}~\cite{colbeck} and by Frank and Lieb~\cite{frank}, \blk repeated here for convenience:
\beq \label{hirrep}
H(A|\rho) + H(B|\rho) \geq S(\rho) + \log_2 C_{AB} .
\eeq
In particular, the additional term $S(\rho)$ on the right hand side represents the inherent uncertainty arising from the state itself.  Unlike Hirschmann's inequality, equation~(\ref{hirrep}) is tight even for position and momentum observables, being saturated in the high-temperature limit of harmonic oscillator thermal states~\cite{frank,hallvol}.  

The main aim of this section is to show that equation~(\ref{hirrep}) can be directly obtained from the Holevo bound, by considering suitable ensembles of signal states. In particular, it corresponds to the Holevo bound for the information that can be extracted by a measurement of $A$ on a uniform ensemble of states related by the group of unitary transformations generated by $B$,  as discussed in sections~\ref{sec:num}--\ref{sec:qp}.
Further, equation~(\ref{hirrep}) is generalised to the case of degenerate observables in section~\ref{sec:deg}) and to energy and time observables in section~\ref{sec:time}.
\blk

\subsection{Number and phase}
\label{sec:num}

Consider first a rigid rotator in two dimensions, with phase $\Phi$ and angular momentum $J_z$. These are mutually unbiased observables, with eigenstates related by $\langle \phi|m\rangle= (2\pi)^{-1/2}e^{im\phi}$.
A rotation of any given state $\rho$ of the rotator, by an angle $\theta$, is generated by applying the unitary transformation
\beq \label{uphi}
\rho_\theta = e^{-iJ_z\theta/\hbar}\rho e^{iJ_z\theta/\hbar} ,
\eeq
as is easily checked via 
\beq \label{shift} p(\phi|\rho_{\theta})=\langle\phi|\rho_\theta|\phi\rangle=\sum_{m,m'}\langle\phi|m\rangle\langle m|\rho_\theta|m'\rangle\langle m'|\phi\rangle=p(\phi-\theta|\rho) .
\eeq
Suppose now that one has a uniform ensemble of such states, $\ce=\{ \rho_\theta;p(\theta)\}$, with a uniform prior probability density $p(\theta)=(2\pi)^{-1}$ , and that the phase observable $\Phi$ is measured on each member of this ensemble. The Holevo bound~(\ref{chi}) for the Shannon information therefore simplifies to
\beq \nn
H(\Phi|\rho_\ce) - \frac{1}{2\pi} \int_0^{2\pi} d\theta\, H(\Phi|\rho_\theta) \leq S(\rho_\ce) -  \frac{1}{2\pi} \int_0^{2\pi} d\theta\, S(\rho_\theta) . 
\eeq
Note that, similarly to equation~(\ref{infbound}), the right hand side is just the $\cg$-asymmetry corresponding to the group of phase shifts of the rotator. Now, $S(\rho_\theta)=S(\rho)$ and $H(\Phi|\rho_\theta)=H(\Phi|\rho)$, from  equations~(\ref{uphi}) and~(\ref{shift}) respectively.  Moreover, the ensemble density operator  $\rho_\ce = (2\pi)^{-1}\int_0^{2\pi}d\theta\,e^{-iJ_z\theta/\hbar}\rho e^{iJ_z\theta/\hbar}$ is clearly invariant under rotation and with the same statistics for $J_z$ as $\rho$, implying the diagonal form
\beq
\rho_\ce =\sum_m |m\rangle\langle m| \,\langle m|\rho|m\rangle
\eeq
%given by
%\begin{align}
%\rho_\ce &= \int_0^{2\pi}\frac{d\theta}{2\pi}\,e^{-iJ_z\theta/\hbar}\rho e^{iJ_z\theta/\hbar} = \sum_{m,m'} \langle m|\rho|m'\rangle \int_0^{2\pi}\frac{d\theta}{2\pi}\,e^{-iJ_z\theta/\hbar}|m\rangle\langle m'| e^{iJ_z\theta/\hbar} \nn\\
%&=\sum_{m,m'} |m\rangle\langle m'|\,\langle m|\rho|m'\rangle\int_0^{2\pi}\frac{d\theta}{2\pi}\,e^{-i(m-m')\theta} = \sum_m |m\rangle\langle m| \,\langle m|\rho|m\rangle ,
%\end{align}
analogously to equation~(\ref{rphi}). Hence, $p(\phi|\rho_\ce)= \sum_m |\langle m|\phi\rangle|^2\langle m|\rho|m\rangle  = (2\pi)^{-1}$ and $S(\rho_\ce) = H(J_z|\rho)$. Substituting into the above Holevo bound and rearranging then gives the entropic uncertainty relation
\beq \label{jphi}
H(J_z|\rho) + H(\Phi|\rho) \geq  S(\rho) +\log_2 2\pi ,
\eeq
corresponding to the strong entropic uncertainty relation Eq.~(\ref{hirgen}) for phase and angular momentum, as desired. 

The same relation holds if $J_z$ is reinterpreted as the spin component of a spin-$j$ particle.  In particular, the Hilbert space of such a particle is spanned by the $2j+1$  eigenstates $\{|j,m\rangle\}$, and hence is in one-one correspondence with the Hilbert space spanned by rotator states $\{|m\rangle:|m|\leq j\}$.

A similar strong uncertainty relation holds for the optical phase $\Phi$ and photon number $N$ observables of a single-mode optical field, as this case is formally equivalent to a rotator with states restricted to have support on nonnegative values of angular momentum, corresponding to replacing $J_z=\hbar\sum_{m=-\infty}^\infty m|m\rangle\langle m|$ by $N=\sum_{n=0}^\infty n|n\rangle\langle n|$. The phase kets are now non-orthogonal, but still satisfy the maximum incompatibility property $\langle\phi|n\rangle=(2\pi)^{-1/2}e^{in\phi}$ (and hence the completeness property $\int_0^{2\pi}d\phi\,|\phi\rangle\langle\phi|=\sum_{n=0}^\infty |n\rangle\langle n| =\id$)~\cite{helstrom,holevo,busch}. Thus, evaluating the Holevo bound for a uniform ensemble of phase-shifted states $\ce_\Phi=\{e^{-iN\theta}\rho e^{iN\theta};(2\pi)^{-1}\}$ leads to the corresponding strong entropic uncertainty relation
\beq \label{nphi}
H(N|\rho) + H(\Phi|\rho) \geq  S(\rho)+ \log_2 2\pi 
\eeq
for number and phase. Note this is equivalent to the geometric form
\beq \label{phiuncert}
\frac{L_\Phi}{2\pi} \geq \frac{V(\rho)}{V_N} \geq \frac{1}{V_N},
\eeq
with $L_\Phi$ denoting the ensemble length $2^{H(\Phi|\rho)}$ of the phase probability density, and $V_N$ denoting the volume of the photon number distribution. Thus, {\it the fractional uncertainty in phase, as measured by its ensemble length,  is never less than the volume of the probe state divided by the spread of the photon number distribution}. It is of interest to compare this uncertainty relation with equation~(\ref{phasefrac}) for general phase estimation.

%Finally, using the variational inequality~(\ref{mow}), the corresponding relation in equation~(\ref{eps}) for continuous variables, and equation~(\ref{phiuncert}), yields the weaker Heisenberg-type uncertainty relation
%\beq
%{\rm Var}\, \Phi\, [{\rm Var}\, N+1/12] \geq \frac{V(\rho)^2}{e^2} = \frac{2^{2S(\rho)}}{e^2} .
%\eeq
%A similar uncertainty relation may be obtained for the rigid rotator, with $N$ replaced by $J_z/\hbar$.

\subsection{Mutually unbiased observables in finite dimensions}

Two mutually unbiased observables $A$ and $B$ on a $d$-dimensional Hilbert space have basis states satisfying $|\langle a|b\rangle|^2=d^{-1}$, and hence $C_{AB}=d$. It is convenient to choose the eigenvalue range $a,b\in\{0,1,\dots,d-1\}$. Further, for the purpose of obtaining uncertainty relations, the eigenstates of $B$ can be rephased without loss of generality such that $\langle b|a\rangle=d^{-1/2} e^{-2\pi iab/d}$ (since such rephasing does not change the measurement distributions of $A$ and $B$).  It follows that 
\beq \label{trans}
e^{-2\pi ijB/d} |a\rangle = \sum_b e^{-2\pi ijB/d} |b\rangle\langle b|a\rangle = d^{-1/2}\sum_b e^{-2\pi i(a+j)b/d}|b\rangle = |a\oplus j\rangle ,
\eeq
where $\oplus$ denotes addition modulo $d$, i.e., $B$ generates translations of $A$ (and vice versa, of course).

Similarly to the previous subsection, consider now a uniform ensemble of translated signal states $\ce=\{\rho_j;p_j\}$, with $\rho_j=e^{-2\pi ijB/d}\rho e^{2\pi ijB/d}$, $p_j=d^{-1}$,  and ensemble density operator $\rho_\ce = \sum_{b,b'} |b\rangle\langle b| \langle b|\rho|b\rangle$. \blk If $A$ is measured on each state, the Holevo bound~(\ref{chi}) then reduces to
\[
H(A|\rho_\ce) - d^{-1} \sum_{j=0}^{d-1} H(A|\rho_j) \leq \log_2 d -  S(\rho) ,
\]
%\begin{align*}
%\rho_\ce &= d^{-1}\sum_{j=0}^{d-1} e^{-2\pi ijB/d}\rho e^{2\pi ijB/d} = \sum_{b,b'} \langle b|\rho|b'\rangle d^{-1}\sum_{j=0}^{d-1} e^{-2\pi ij(b-b')/d}|b\rangle\langle b'| \\
%& = \sum_{b,b'} |b\rangle\langle b| \langle b|\rho|b\rangle .
%\end{align*}
%This is diagonal in $B$, yielding $S(\rho_\ce)=H(B|\rho)$ and $p(a|\rho_\ce)=d^{-1}$. Moreover, from equation~(\ref{trans}) we have $S(\rho_j)=S(\rho)$ and $p(a|\rho_j)=\langle a|\rho_j|a\rangle=p(a \ominus j|\rho)$. 
 and using Eq.~(\ref{trans}) then \blk yields the strong entropic uncertainty relation
\beq \label{ab}
H(A|\rho) + H(B|\rho) \geq S(\rho) + \log d,
\eeq
as required. Note that, using equation~(\ref{vol}), this may be written in the geometric form
\beq \label{fracform}
\frac{V_A}{d}\,\, \frac{V_B}{d} \geq \frac{V(\rho)}{d}
\eeq
for the fractional volumes occupied by the corresponding classical and quantum ensembles.

\subsection{Position and momentum}
\label{sec:qp}

The case of position and momentum observables is a little less straightforward, essentially because there is no uniform probability measure over the set of signal states corresponding to all possible translations of $\rho$.  However, this can be dealt with by taking a suitable limit of ensembles of such states. In particular, the Fourier transform relation connecting position and momentum eigenkets $\{|q\rangle\}$ and $\{|p\rangle\}$ can be regarded as the limit of a discrete Fourier transform on a finite Hilbert space, allowing the desired entropic uncertainty relation to be obtained via equation~(\ref{ab}) for discrete observables.  Thus the `hard work' has already been done in the previous section.

A simple approach to taking suitable limits, for the case of a particle moving in one dimension, is to approximate $Q$ and $P$ by  two maximally incompatible observables $Q_d$ and $P_d$ on a Hilbert space subspace of $d=2r+1$ dimensions, with spectral decompositions
\beq
Q_d = \sum_{m=-r}^r q_m |\psi_m\rangle\langle \psi_m|,\qquad\qquad
P_d = \sum_{n=-r}^r p_n |\phi_n\rangle\langle \phi_n| .
\eeq
Here the eigenvalues are chosen to correspond to equally spaced position and momentum values 
\beq \label{deltaqp}
 q_m= m\,\delta q := m\frac{L}{\sqrt{d}}, \qquad\qquad p_n=n\,\delta p := n \frac{2\pi\hbar}{L\sqrt{d}},
\eeq
with spacings $\delta q$ and $\delta p$ respectively defined via an arbitrary  fixed length $L$, and the eigenstates by
\beq \nn
\psi_m(q):=\left\{ \begin{array}{cc} 1/\sqrt{\delta q},& |q-q_m|\leq \half\delta q\\ 0,& {\rm otherwise}\end{array} \right. ,~~~
|\phi_n\rangle := d^{-1/2} \sum_m e^{2\pi imn/d} |\psi_m\rangle .
\eeq
Note that $\delta q, \delta p\rightarrow 0$ as $d\rightarrow\infty$, and that $\delta q\delta p = 2\pi\hbar/d$.
Moreover, from equation~(\ref{ab}) of the previous section we immediately have the entropic uncertainty relation
\beq \label{diseur}
H(Q_d|\rho_d) + H(P_d|\rho_d) \geq S(\rho_d) + \log_2 d,
\eeq
for any state $\rho_d$ with support in the subspace spanned by $\{|\psi_m\rangle\}$.  

To show explicitly that $Q_d$ and $P_d$ are indeed approximations of $Q$ and $P$,  note first that as $d\rightarrow\infty$, i.e., as $\delta q\rightarrow 0$,
\begin{align} 
p(Q_d=q_m|\rho_d) = \langle \psi_m|\rho_d|\psi_m\rangle  &= \frac{1}{\delta q} \int_{q_m-\delta q/2}^{q_m+\delta q/2} dq\int_{q_m-\delta q/2}^{q_m+\delta q/2} dq'\,\langle q|\rho_d|q'\rangle \nn \\
&\approx \delta q \,\langle q_m|\rho_d|q_m\rangle = \delta q\,p(Q=q_m|\rho_d) .
\end{align}
Hence the distribution of $Q_d$ is a discretised approximation of the distribution of $Q$, implying that as $d\rightarrow \infty$ their entropies are related by
\begin{align}
H(Q_d|\rho_d)&=-\sum_{m=-r}^r p(Q_d=q_m|\rho_d) \log_2 p(Q_d=q_m|\rho_d) \nn\\
&\approx -\sum_{m=-r}^r p(Q_d=q_m|\rho_d) \log_2 [\delta q\,p(Q=q_m|\rho_d)] \nn\\
&\approx -\log_2 \delta q-\sum_{m=-r}^r \delta q \,p(Q=q_m|\rho_d) \log_2  p(Q=q_m|\rho_d) \nn\\
&\approx -\log_2 \delta q-\int_{-\infty}^\infty dq\,p(Q=q_m|\rho_d) \log_2 p(Q=q_m|\rho_d)\nn \\
& = H(Q|\rho_d) - \log_2 \delta q.  \label{hqd}
\end{align}
Similarly, the distributions of $P_d$ and $P$ are related in the limit $d\rightarrow\infty$ via
\begin{align}
p(P_d=&p_n|\rho_d) = \langle \phi_n|\rho_d|\phi_n\rangle
= \sum_{m,m'=-r}^r\langle \phi_n|\psi_m\rangle \langle\psi_m|\rho_d|\psi_{m'}\rangle\langle \psi_{m'}|\phi_n\rangle \nn\\
&= \frac{1}{d\delta q}\sum_{m,m'=-r}^r e^{-2\pi i(m-m')n/d}  \int_{q_m-\delta q/2}^{q_m+\delta q/2} dq\int_{q_{m'}-\delta q/2}^{q_{m'}+\delta q/2} dq'\,\langle q|\rho_d|q'\rangle \nn \\ 
&\approx \frac{\delta q}{d} \sum_{m,m'=-r}^r e^{-ip_n (q_m-q_{m'})/\hbar} \langle q_m|\rho_d|q_{m'}\rangle \nn \\ 
&=\delta p \left(\sum_{m=-r}^r \frac{\delta q  e^{-ip_n q_m/\hbar}}{\sqrt{2\pi\hbar}} \langle q_m|\right) \rho_d \left(\sum_{m=-r}^r \frac{\delta q e^{ip_n q_{m'}/\hbar}}{\sqrt{2\pi\hbar}}  | q_{m'}\rangle\right) \nn\\
&\approx \delta p \,\langle p_n|\rho_d|p_n\rangle = \delta p \,p(P=p_n|\rho_d) .
\end{align}
Thus, the former distribution is a discretised approximation of the latter, implying as $d\rightarrow\infty$ that
\beq \label{hpd}
H(P_d|\rho_d) \approx  H(P|\rho_d) - \log_2 \delta p ,
\eeq
similarly to equation~(\ref{hqd}) above. Adding equations~(\ref{hqd}) and~(\ref{hpd}) gives
\[
H(Q_d|\rho_d) + H(P_d|\rho_d) \approx H(Q|\rho_d)+H(P|\rho_d) -\log_2 (\delta q\delta p). \]
Substituting this into uncertainty relation~(\ref{diseur}) 
for $\rho_d:=E\rho E/\tr{E\rho E}$, where $E=\sum_m |\psi_m\rangle\langle\psi_m|$ is the projection onto the $d$-dimensional subspace, and taking the limit as $d\rightarrow\infty$, then yields the strong entropic uncertainty relation
\beq
H(Q|\rho) + H(P|\rho) \geq S(\rho) + \log_2 2\pi\hbar
\eeq
as desired (when the corresponding classical and quantum entropies exist).

The above elementary construction may be easily generalised to the case of particles moving in $D$  dimensions, via replacement of the one-dimensional intervals of lengths $\delta q$ and $\delta p$ in equation~(\ref{deltaqp}) by $D$-dimensional cells with volumes $(\delta q)^D$ and $(\delta p)^D$ respectively, leading to
\beq
H(Q|\rho) + H(P|\rho) \geq S(\rho) + D \log_2 2\pi\hbar .
\eeq
Note that this is equivalent to the geometric form given in equation~(\ref{xpgeo}).

\subsection{Degenerate observables}
\label{sec:deg}

If $\h_{AB}$ denotes the Hilbert space spanned by the eigenkets $|a\rangle$ and $|b\rangle$, consider now a quantum system on a larger Hilbert space $\h=\h_{AB}\otimes\h_Z$. The original Hirschmann inequality~(\ref{hir}) still holds for states $\rho$ on this Hilbert space, since one can replace $\rho$ by
${\rm tr}_Z[\rho]$ on $\h_{AB}$ without changing the entropies of $A$ and $B$.  However, inequality~({\ref{hirrep}) is modified by such a replacement, to
\beq \label{hirmod}
H(A|\rho) + H(B|\rho) \geq S(\tilde\rho) + \log_2 C_{AB} ,\qquad \tilde\rho:= {\rm tr}_Z[\rho] 
\eeq 
for mutually unbiased observables of a subsystem embedded in a larger system. 

An alternative uncertainty relation for such observables can be obtained by applying the Holevo information bound technique directly to the larger Hilbert space.  For example, for the case of phase and photon number, the first line of equation~({\ref{phasesingle}) implies that this technique leads to the entropic uncertainty relation
\begin{align}
H(N|\rho) + H(\Phi|\rho) &\geq \log_2 2\pi + S(\rho)- \sum_n p_n S(\Pi_n\rho \Pi_n/p_n) \nn\\
&= \log_2 2\pi + S(\rho)- \sum_n p_n S(|n\rangle\langle n|\otimes \rho_{a|n}) \nn\\ \label{degen}
&= \log_2 2\pi + S(\rho)- \sum_n p_n S(\rho_{a|n}),
\end{align}
where $\rho_{a|n}$ is the state of the auxiliary system corresponding to outcome $N=n$. Note that this is equivalent to equation~(\ref{hirmod}) if $\rho$ is pure or if $\rho=\tilde\rho\otimes \rho_a$. This approach is exemplified below for the  case of the energy and time observables of a free particle.

\subsection{Energy and time}
\label{sec:time}

\subsubsection{Free particle}

A quantum system with Hamiltonian $E$ and a continuum of energy eigenstates $\{|\epsilon,d\rangle\}$, where $\epsilon\in [0,\infty)$ is the energy eigenvalue and $d$ labels any degeneracies, has a canonical time observable $T$ corresponding to the POVM  $\{T_t=\sum_d|t,d\rangle\langle t,d|\}$, where $|t,d\rangle:= \int d\epsilon\,(2\pi\hbar)^{-1/2}e^{-i\epsilon t/\hbar}|\epsilon,d\rangle$~\cite{holevo,halltime,busch}.  These states are typically non-orthogonal, and so $T$ cannot be represented by a Hermitian operator in general. \blk

In the case of no degeneracies $E$ and $T$ are mutually unbiased, with $C_{ET}=2\pi\hbar$, and so are formally equivalent to conjugate momentum and position observables (for states restricted to positive momentum eigenvalues). 
However, such systems are typically degenerate, and the generalisations in
 the previous subsection are required to obtain entropic uncertainty relations. 
 
 Here the example of a free particle of mass $m$ moving in one dimension will be considered, with Hamiltonian $E=(2m)^{-1}P^2$. The energy eigenstates may be labelled by $|\epsilon,\pm\rangle$, with the sign corresponding to eigenstates of positive and negative momentum, respectively. Hence any state $\rho$ of the particle may be decomposed as
\beq
\rho = \sum_{\alpha,\beta=\pm} \int d\epsilon\,d\epsilon' |\epsilon,\alpha\rangle\, \langle \epsilon',\beta|\, \langle\epsilon,\alpha|\rho|\epsilon',\beta\rangle  .
\eeq
Note that the Hilbert space is isomorphic to the tensor product of an infinite-dimensional and  two-dimensional Hilbert space, i.e., $\h\equiv \h_{ET}\otimes \h_Z$, where $\h_Z={\rm span}\{|+\rangle,|-\rangle\}$ corresponds to the degenerate component of the energy. This corresponds to the identification $|\epsilon,\alpha\rangle\equiv |\epsilon\rangle\otimes|\alpha\rangle$. Applying the Holevo bound approach used in equation~(\ref{degen}) then yields the entropic uncertainty relation
\beq \label{free}
H(E|\rho) + H(T|\rho) \geq \log_2 2\pi\hbar + S(\rho) - \int d\epsilon\, p(\epsilon|\rho) S(\rho_{a|\epsilon})
\eeq
where $\rho_{a|\epsilon}$ is the conditional state on $\h_Z$ corresponding to $E=\epsilon$, i.e.,
\beq 
\rho_{a|\epsilon} = \sum_{\alpha,\beta} |\alpha\rangle\langle\beta|\,\frac{\langle \epsilon,\alpha|\rho|\epsilon,\beta\rangle}{p(\epsilon|\rho)} .
\eeq

For the case that $\rho$ is pure, the last two terms in equation~(\ref{free}) vanish, and the relation is analogous to the Hirschmann inequality~\ref{hir}) for position and momentum. Further, in the case that $\rho=\tilde\rho\otimes\rho_a$ the relation reduces to
\beq
H(E|\rho) + H(T|\rho) \geq \log_2 2\pi\hbar + S(\tilde\rho) .
\eeq
This generalises Grabowski's uncertainty relation for the case of a pure state having support only on positive momentum eigenstates~\cite{grab}.  It would be of interest to calculate the right hand side of equation~(\ref{free}) for the case of a Gaussian mixed state.

\subsubsection{Harmonic oscillator}

A one-dimensional harmonic oscillator of frequency $\omega$ has Hamiltonian $E=\hbar\omega N$, proportional to the number operator, and hence the canonical time observable $T$ is similarly proportional to the phase $\Phi$ of the oscillator, i.e., $\Phi\equiv\omega T$. %(with corresponding POVM is $\{|t\rangle\langle t|, t\in [0,2\pi\omega^{-1}\}$, with $\langle t|n\rangle=(2\e^{-in\omega t}$). 
It immediately follows that $p(t|\rho)=\omega\, p(\phi|t)$, and hence that
\beq
H(T|\rho)=H(\Phi|\rho)- \log_2 \omega = H(\Phi|\rho) +\log_ 2\pi -\log\tau,
\eeq
where $\tau$ is the period of the oscillator.  Noting that $H(E|\rho)=H(N|\rho)$, substitution into equation~(\ref{nphi}) for number and phase observables then gives the corresponding simple generalisation
\beq \label{per}
H(E|\rho) + H(T|\rho) \geq S(\rho)+ \log_2 \tau 
\eeq
for the energy and time observables of the oscillator.

\subsubsection{Systems with discrete energy levels}

Finally, consider the case of a nondegenerate Hamiltonian $E$ with discrete eigenvalues $\{\epsilon_n\}$. The canonical time observable in this case has been discussed in \cite{halltime,hallpegg}, including a generalisation of the Hirschmann inequality~(\ref{hir}) for energy and time~\cite{halltime}, and further details may be found in these references.

Such a system is almost-periodic, and can be approximated arbitrarily well for arbitrarily long times by a periodic system having a sufficiently long period.  Now, for a periodic system with period $\tau$, the time probability density $p(t|\rho)$ on $[0,\tau)$ can be extended to a periodic function on the real line, via $p(t+\tau|\rho):= p(t|\rho)$.  Defining the related function on the real line via $p_{ap}(t|\rho)=\tau\, p(t|\rho)$, one trivially has the positivity and normalisation properties
\beq
p_{ap}(t|\rho) \geq 0,\qquad \lim_{x\rightarrow\infty} \frac{1}{2x}\int_{-x}^x dt\, p_{ap}(t|\rho) =  1 .
\eeq
Any function with these properties is called an almost-periodic probability density, and the canonical time observable of any almost-periodic system is described by such a function~\cite{halltime,hallpegg}. 

The corresponding almost-periodic entropy  is defined by~\cite{halltime}
\beq
H_{ap}(T|\rho) := -\lim_{x\rightarrow\infty} \frac{1}{2x}\int_{-x}^x dt\, p_{ap}(t|\rho) \log_2 p_{ap}(t|\rho) .
\eeq
For the periodic case $p_{ap}(t|\rho)=\tau p(t|\rho)$ this simplifies to $H_{ap}(T|\rho) = H(T|\rho) - \log_2 \tau$, and hence the uncertainty relation~(\ref{per}) for periodic systems can be rewritten as
\beq 
H(E|\rho) + H_{ap}(T|\rho) \geq S(\rho) , 
\eeq
independently of $\tau$.  Finally, since any almost-periodic quantum system can be approximated arbitrarily well by a periodic system of sufficently large period, as noted above, this entropic uncertainty relation also holds for the general case of almost-periodic systems.

\section{Conclusions}
\label{conc}

The Holevo information bound provides a general connection between classical and quantum entropies.  It is this precisely this connection that allows it to be used to derive the Heisenberg limits and entropic uncertainty relations in sections~\ref{heislim} and~\ref{eurs}, which similarly feature both classical and quantum entropies.

Entropic Heisenberg limits, such as equations~(\ref{entdiff}) and~(\ref{heisag}), bound the decrease in uncertainty, of an unknown group displacement of a probe state, by the $\cg$-asymmetry of the state (and also by the photon number entropy of the state in the case of phase displacements).  For the special case of covariant estimates, as noted following equation~(\ref{lower}), this decrease in uncertainty is precisely the average information gain per probe state.

Whereas entropy is used as a measure of uncertainty in the above limits, a more direct measure is its exponential, which has the geometric properties of a `volume' (or a `length' for one-dimensional probability distributions). This leads to corresponding geometric Heisenberg limits, such as equations~(\ref{phasefrac}), (\ref{genlim}) and~(\ref{fundamental}), for the ratio of initial and final uncertainties,  as characterised by the volumes of the prior probability density and the error probability density. \blk Further, the lower bounds are themselves  volume ratios, e.g., of the probe state and its average over random displacements.  These bounds  may also be used to derive strong Heisenberg limits for root mean square error and related quantities, including in terms of average photon number as per equation~(\ref{epsheis}), and for multiparameter displacements such as rotations induced by a magnetic field as per Eqs.~(\ref{something}-\ref{terr}).

Similarly, the strong entropic uncertainty relations derived in section~(\ref{eurs}), such as equations~(\ref{xpgeo}), (\ref{phiuncert}) and~(\ref{fracform}), for position and momentum, number and phase, and finite Hilbert spaces, respectively, may be interpreted geometrically in terms of the volumes of classical and quantum ensembles. The derivations are particularly simple in that they arise directly from the Holevo bound, when applied to a random ensemble of suitably displaced states, and generalise straightforwardly to relations for degenerate observables and for energy and time observables.

A number of possibilities for future work on entropic and geometric Heisenberg limits have already been noted in section~\ref{heislim}, mostly relating to calculation of limits for multicomponent probe states (such as $M$ copies of a given state). It would also be of interest to extend results to discrete groups, and to noncompact groups of displacements such as position displacements of a quantum probe. In the latter case, the uniform probability density $p(g)$ in the randomisation operation $\R_\cg$ in equation~(\ref{rg}) would need to be replaced by a density with a finite entropy and volume. For example, the choice of a Gaussian density for $p(g)$, applied to Gaussian probe states, is expected to be tractable. Finally, while the control error $g_{err}$ in equation~(\ref{gerr}) has been proposed as a suitable `error' variable in the general case, one could also consider bounds for the average fidelity between the displaced and estimated probe state.

There are also a number of possibilities for generalising the derivation entropic uncertainty relations from Holevo's bound to obtain other useful entropic inequalities.  For example, if $\Phi_{cov}$ is any covariant phase observable, then precisely the same argument used to derive equation~(\ref{jphi}) may also be used to derive the entropic uncertainty relation
\beq 
H(J|\rho) + H(\Phi_{cov}|\rho) \geq  S(\rho)+ \log_2 2\pi .
\eeq
It may also be of interest to consider entropic uncertainty relations for Gaussian measurements on Gaussian states, by calculating the corresponding Holevo bounds.

Finally, ideas from sections~\ref{heislim}) and~(\ref{eurs}) can be combined in various ways to obtain lower bounds for the entropies of observables.  For example, consider a particle moving in three dimensions, labelled by $(r, \theta,\phi)$ in spherical coordinates. The corresponding Holevo bound $A_\cg(\rho)$, for a measurement of the angular variables $\Theta$ and $\Phi$ on the ensemble of random rotations of the state of the particle, follows via equation~(\ref{aggen}) as
\beq 
\log_2 4\pi - H(\Theta,\Phi|\rho) \leq H(J^2|\rho) + \langle \log_2(2j+1)\rangle-S(\rho). 
\eeq
While this is not a particularly useful entropic uncertainty relation {\it per se}, it can be used to bound the uncertainty of $\Theta$ and $\Phi$ in terms of the angular momentum properties of the state.  For example, for a particle with fixed total angular momentum $J^2=j(j+1)\hbar^2$ it follows that
\beq
A_{\Theta,\Phi} \geq \frac{4\pi}{2j+1} V(\rho) \geq \frac{4\pi}{2j+1} ,
\eeq
for the effective spherical area $A_{\Theta,\Phi}=2^{H(\theta,\Phi|\rho)}$ occupied by the probability density $p(\theta,\phi|\rho)$. A similar bound is obtained for the case of a particle with $j\leq j_{max}$, using equation~(\ref{agjmax}), with $2j+1$ in the above equation replaced by $(j_{max}+1)^2$.

\flushleft{\bf Acknowledgements}  
I thank Dominic Berry for helpful discussions on multimode phase estimation.

%\bibliographystyle{iopart-num}

%\bibliographystyle{unsrt}

%\bibliography{holevo}

\end{document}